\documentclass{aastex}
\usepackage{emulateapj5}
\begin{document}
\title{The Dark Halo of NGC 5963 as a Constraint on Dark Matter Self-Interaction
at the Low Velocity Regime }
\author{F.J. S\'{a}nchez-Salcedo }
\affil{Instituto de Astronom\'\i a, UNAM, Ciudad Universitaria, Aptdo. 70 264,
C.P. 04510, Mexico City, Mexico}
\email{jsanchez@astroscu.unam.mx}

\begin{abstract}
Self-interacting dark matter has been proposed as a hypothesis
to explain the shallow central slopes of the density profiles
of dark matter halos in galaxies. In order to be consistent with
observational studies at scales of galaxy clusters, the cross section
should scale inversely with the velocity of collision.
In this paper we consider the mass density profile of the halo
of the low surface brightness (LSB) galaxy NGC 5963 to place an upper
limit on the dark matter cross section for collisions with velocities
$\sim 150$ km s$^{-1}$, i.e.~at the low velocity regime.
After calibrating against cosmological simulations, we found that
the large inferred dark matter concentration and central dark matter
density in NGC 5963 are inconsistent with an effective collisional 
cross section per unit of mass $>0.2$ cm$^{2}$ g$^{-1}$. 
Corrections were applied in order to account for reduction of the core
by the adiabatic contraction caused by cooling baryons.
Our limits that involve a number of simplifying, but always
conservative, assumptions, exclude the last permitted interval for
velocity-dependent cross sections to explain 
the flat density core in LSB galaxies. Implications for the nature of 
dark matter are also discussed.



\end{abstract}
\section{Introduction}
Numerical studies of structure formation with the collisionless 
cold dark matter (CDM) scenario predict
dark halos with steep central cusps (e.g., Navarro, Frenk \& White 1996, 1997,
hereafter NFW), whereas most of the rotation 
curves of dwarf galaxies and low surface brightness (LSB)
galaxies suggest that their halos
have constant density cores (e.g., Marchesini et al.~2002;
de Blok \& Bosma 2002, and references
therein). Self-interacting dark matter (SIDM) with cross sections 
per unit of mass in the range $0.5$--$6$ cm$^{2}$ g$^{-1}$ was proposed 
as a possible route to reduce the
density cusp, as thermal conduction replaces the central cusp by a 
soft core (Spergel \& Steinhardt 2000). 
A tight constraint on the cross section just under $0.1$
cm$^{2}$ g$^{-1}$ has been derived by considering the formation
of giant cluster arcs (Meneghetti et al.~2001) and the size of
the cores of galaxy clusters (e.g., Arabadjis, Bautz
\& Garmire 2002; Lewis, Buote \& Stocke 2003; Arabadjis \& Bautz 2004).
Since this cross section is too small to produce galactic cores,
it has been pointed out that a velocity dependent cross section 
might reduce the effects of self-interaction on cluster scales
(e.g., Firmani et al.~2000). For a cross section per unit of mass, 
which varies as some power of the relative velocity between colliding 
particles $\sigma_{\rm dm}=\sigma_{\ast}(v_{0}/v_{\rm rel})^{a}$,
where $v_{0}=(100\, {\rm km \,s}^{-1})$, there exists a certain range
of values of $\sigma_{\ast}$ and $a$ for which the halos of dwarf
galaxies should present a core, while halos of galaxy clusters
would not change their core sizes, ellipticities and arcs significantly.



Constraints on the parameters $\sigma_{\ast}$ and $a$ were obtained
by requiring that dwarf galaxies observed today have yet to undergo
core collapse and that dark halos must survive the heating from
hot cluster halos (Gnedin \& Ostriker 2001; Hennawi \& Ostriker 2002). 
All these requirements
are satisfied for a very narrow range of parameters $\sigma_{\ast}=0.5$--$1$
cm$^{2}$ g$^{-1}$ and $a=0.5$--$1$. For cross sections 
within this suitable range, flat cores with
densities of $\sim 0.02$ M$_{\odot}$ pc$^{-3}$ are formed in
the central regions of galactic halos as it was confirmed numerically
by Col\'{\i}n et al.~(2002) in cosmological simulations of SIDM.

Low surface brightness galaxies are ideal to put upper limits
on the strength of dark matter (DM) self-interaction at the low-velocity
regime (i.e.~relative velocities of $\sim 150$ km s$^{-1}$). 
In contrast to relatively
recently formed objects, like clusters of galaxies, that present
unrelaxed mass distributions, LSB galaxies
with high central densities may have the highest formation
redshifts and, therefore, they have had almost a Hubble time to 
soften their cusps.
In the present study we concentrate on the implications of the halo
of the LSB galaxy NGC 5963 for dark matter self-interaction. The
high central density and the small core radius of the halo of NGC 5963
suggest that
either the DM cross section, at the low velocity regime,
is rather small $\lesssim 0.1$
cm$^{2}$ g$^{-1}$ or the dark halo is undergoing an undesirable,
dramatic core collapse.
Therefore, it seems
very unlikely that collisional scattering between DM particles 
is the main agent for the formation of cores.

We will start in \S \ref{sec:basic} with a description of the
basic physics behind the scenario of SIDM and some predictions, 
which will be used in the remainder of the paper.
In \S \ref{sec:thegalaxy} we briefly describe the properties of NGC 5963 
and present mass models over a range of mass-to-light ratios of the 
stellar disk.
We then compare with predictions of cosmological SIDM simulations to
constrain the self-interaction cross section of DM particles 
(\S \ref{sec:constraints}).
Some implications of the results are discussed in \S \ref{sec:discussion}.

\section{Self-Interacting Dark Matter and the Size of the Cores}
\label{sec:basic}
The mass-density profiles of numerically simulated collisionless CDM halos
are commonly parameterized with the analytical Navarro, Frenk \& White 
profile:
\begin{equation}
\rho (r)=\frac{\rho_{i}}{\left(r/r_{s}\right)
\left(1+r/r_{s}\right)^{2}},
\label{eq:NFW}
\end{equation}
where $r_{s}$ is the characteristic radius of the halo and $\rho_{i}$
is related to the density of the Universe at the time of collapse. 
Within the core region, the dark matter density increases as a power law
$\rho\sim r^{-1}$ and the velocity dispersion decreases toward the center
at the radius $r_{s}$.
The parameters $(\rho_{i},r_{s})$ are usually expressed in a slightly
different form by the concentration parameter $c=r_{200}/r_{s}$ and the
rotation velocity $V_{200}$ at radius $r_{200}$. The latter is the
radius inside which the average overdensity is $200$ times the critical
density of the Universe.

If dark matter particles experience self-interaction, the scattering
thermalizes the inner regions of dark halos, producing a constant-density
core (e.g., Burkert 2000). It is expected that the core radius
will be comparable to the inversion radius, i.e.~the radius
at which the velocity dispersion peaks. After reaching the maximum core size, 
the core can begin to recollapse due to the gravothermal catastrophe. 
In order to have a core, the relaxation time should be less than
$\sim 1/10$ of Hubble time, so that it should have sufficient time
to form a core by thermalization, while the lifetime of its core
should exceed $10$ Hubble times, so the core should be far from
collapse. 

In the monolithic scenario in which the halos evolve at isolation, 
the core radius increases in time
as the result of self-interaction between halo particles.
Starting with a profile with inversion radius $r_{i}$, 
the core radius of the pseudoisothermal profile,
$r_{c}$, achieves a radius $\approx 0.4 r_{i}$ in one core-radius relaxation
timescale $t_{\rm rc}=(3\rho\sigma_{\rm dm}
\tilde{v})^{-1}$, with $\rho$ the density and $\tilde{v}$ the velocity
dispersion, both evaluated at  
$0.6 r_{i}$ (e.g., Kochanek \& White 2000)\footnote{Kochanek \& White (2000)
estimated the core radius by the point at which the density drops
to $1/4$ the central density. To estimate $r_{c}$, we have used
that $r_{c}$ is related to
$r_{1/4}$ by the relation $r_{c}=r_{1/4}/1.7$.}.
The maximum radius that the core reaches before recollapse may be 
even larger. In fact,
for the particular case of a Hernquist (1990) model with break radius $r_{H}$,
Kochanek \& White (2000) found that $r_{c}$ reaches a
maximum value of $0.23 r_{H}$.
Since the velocity dispersion for a Hernquist profile peaks at $0.33 r_{H}$,
the core radius $r_{c}$ at the maximum expansion is approximately $0.7 r_{i}$.


The assumptions behind the monolithic scenario that the initial
configuration follows a NFW profile and that halos evolve at 
isolation appear to be inconsistent with the SIDM 
hypothesis in a cosmological context. Cosmological simulations of the formation 
of SIDM halos, either with a constant cross section or dependent on
velocity, have been carried out by Yoshida et al.~(2000),
Dav\'e et al.~(2001), Col\'{\i}n et al.~(2002) and D'Onghia,
Firmani \& Chincarini (2003). 
The inclusion of dynamically
hot material that continuously is accreted onto the halo (in
a cosmological context) can prevent
efficient heat transfer from the core to the halo. Moreover,
it turned out that since collisions already modify halo profiles
in the dynamical process of halo growth, there is a smooth trend of 
increasing core radius with the effective value of $\sigma_{\rm dm}$. 
A core with a radius $\sim 0.4 r_{s}$ 
is formed if the mean number 
of collisions per particle and per
Hubble time in the center halo is about $4$--$6$. 
We will use this result in \S \ref{sec:constraints}. 

The existence of galaxies with large DM concentrations and central
DM densities implies upper limits on the number of collisions
per particle and hence on $\sigma_{\ast}$.
In the following we consider the implications of
the concentrated halo in the galaxy NGC 5963.

\section{The Galaxy NGC 5963 and its Dark Halo}
\label{sec:thegalaxy}
\subsection{NGC 5963 and its Rotation Curve}

\begin{figure}
\plotone{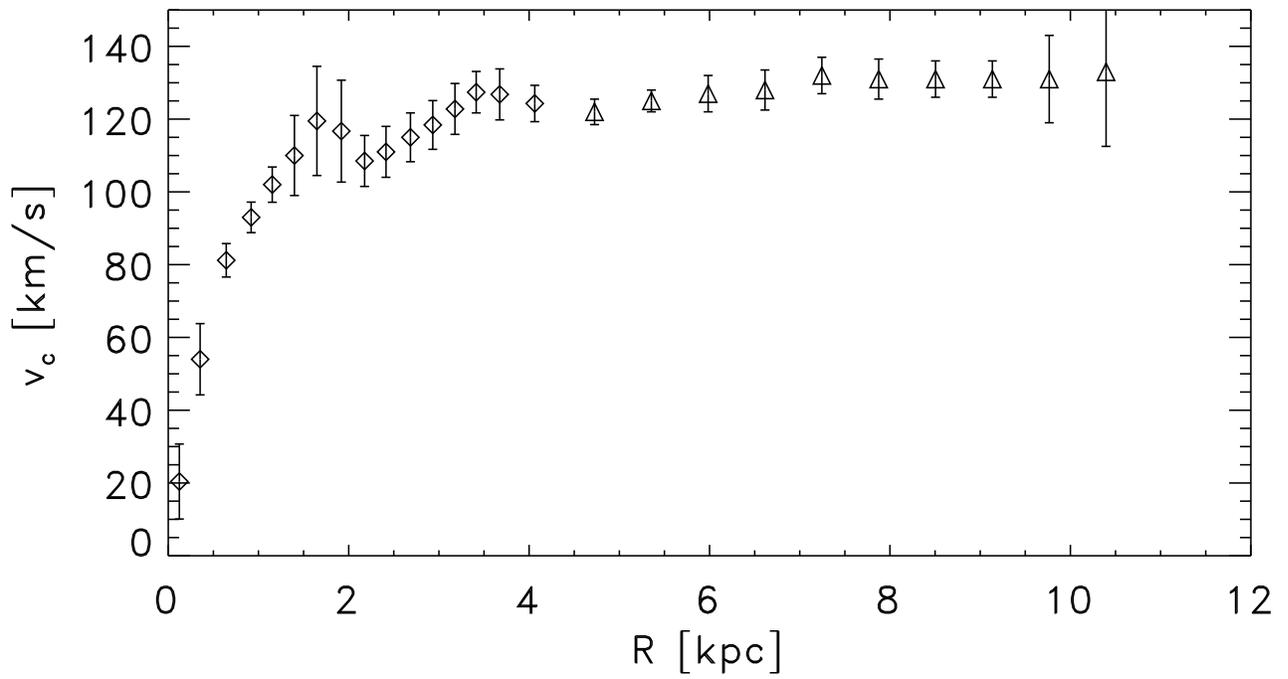}
\caption{
The observed CO rotation curve (empty diamonds) of NGC 5963 taken from 
Bolatto et al.~(2003) together with the H\,{\sc i} rotation
curve (empty triangles) from Bosma et al.~(1988).  }
\label{fig:RC}
\end{figure}

NGC 5963 is a relatively isolated LSB galaxy associated with the
NGC 5866 group, with a suitable inclination angle of $45\pm 4 \deg$. Its  
adopted distance is $\sim 13$ Mpc, based on the recession velocity
derived by Bosma, van der Hulst \& Athanassoula (1988) 
and a Hubble constant of $70$ km s$^{-1}$
Mpc$^{-1}$. The uncertainty in the distance is probably of
order $3$ Mpc. The distributions of light and mass
were first reported by Romanishin et al.~(1982), who found a
luminosity profile that exhibits a lens component, i.e.~a component
with a plateau and a steep outer edge, and a small bulge (see also
Simon et al.~2004). The extent of the lens is
about $1.4$ kpc. The H\,{\sc i} rotation curve was constructed by
Bosma et al.~(1988). These authors emphasized
that the dark halo of this galaxy is more concentrated than in
normal Sc's with similar rotation curves. In this sense, NGC 5963
obeys the rule pointed out by Sancisi (2003) that an excess
of light corresponds an excess of rotation and viceversa.
Beyond a radius of $7$ kpc the
rotation curve becomes fairly flat with a circular velocity $v_{c}$
$\sim 131$ km s$^{-1}$ (see Fig.~\ref{fig:RC}). 
If the dark halo can be approximated 
by the pseudoisothermal sphere, the one-dimensional velocity dispersion
for halo particles for NGC 5963 would be of $\tilde{v}\approx
v_{\rm c}/\sqrt{2}=90$ km s$^{-1}$ and, therefore, the halo
particles undergo collisions in the low-velocity regime.  

In order to study the cuspiness of the mass density profile
in the galaxy centers it is required a rotation curve with a 
high spatial and velocity resolution and a large extension. 
This can be achieved by combining CO, H$\alpha$ data, 
and radio H\,{\sc i} data. 
NGC 5963 is one of the targets of the sample of galaxies with high-resolution
two-dimensional H$\alpha$ and CO velocity fields, which
are free of beam-smearing, in Simon et al.~(2004).
Two-dimensional velocity fields are useful in order to avoid systematic
uncertainties and to determine the existence of radial motions.
Multiple wavelengths studies of the velocity fields may also help to
elucidate the origin of systematic errors that can make density
profiles appear artificially shallow. 
The observations in CO and H$\alpha$ provide the 
rotation curve within a galactocentric radius of $4$ kpc, whereas
the H\,{\sc i} rotation curve allows us to trace the potential out to $10$ kpc.
This galaxy lacks of measurable non-circular motions, i.e.~radial velocities
$< 5$ km s$^{-1}$, except perhaps in the rings between $12''$ to $30''$,
those being not large enough to affect the derivation of the DM
density profile. 
Simon et al.~(2004) compared the H$\alpha$ and CO velocities at every
point and found an excellent agreement between them, with a mean
offset of less than $1$ km s$^{-1}$, and a scatter of $7.8$ km s$^{-1}$.
Though the uncertainties in the rotation curve can be reduced
by combining both velocity data, we prefer to use the CO rotation
curve and its associated errorbars rather than the formal errorbars of the
H$\alpha$ rotation curves, which simply indicate that the Gaussian
fit to the profile was well-determined. In particular, the CO rotation
curve is less sensitive to asymmetric drift corrections and other magnetic 
terms, due to the smaller small-scale motions of the molecular clouds
compared to the ionized gas (S\'anchez-Salcedo \& Reyes-Ruiz 2004). 
In addition, other uncertainties
in mass models, as the mass-to-light ratio of the different components,
impose a minimum error in any fit.
Figure \ref{fig:RC} shows the CO rotation curve in the inner galaxy and the
H\,{\sc i} rotation curve in the outer parts, with error bars indicating
the uncertainties in the observations. 

The present galaxy is an ideal laboratory with which
to test the effects of collisions between DM halo particles
in the low velocity regime, not only for the properties of
the galaxy but also for the quality of the data.

\subsection{Mass Models and Adiabatic Contraction}
We have constructed different mass models for this galaxy.
First we calculate the rotation curve due to the neutral (H\,{\sc i}$+$He)
gas mass from the radial distribution of the H\,{\sc i}
surface density from Bosma et al.~(1988). The 
surface brightness profiles in the optical and near-{\it IR} bands 
given by Simon et al.~(2004)
were used to derive the disk mass distribution
by adopting a constant mass-to-light ratio for the disk.
The resulting mass distribution was converted into a disk rotation
curve. The difference between the sum of the squares of the observed
velocity and the stellar plus gaseous rotation velocities was converted into
a halo mass distribution after assuming a certain form for the halo.
Along this paper, all the fits
to the CO rotation curve were performed by a $\chi^{2}$-minimization. 
To make the best use of both the CO and H\,{\sc i} data, we have
used a hybrid rotation curve. It consists of the CO data over the range
of radii where available and $21$ cm data to define the outermost points.
This was accomplished by minimizing $\chi^{2}$ between $0$ and
$4$ kpc but requiring that between $5$ and $10$ kpc, 
the goodness of the fit must be better than $1\sigma$ level, while rejected
otherwise.
In this way, the fits are statistically consistent with the H\,{\sc i} 
rotation curve at large radii.

Different mass decompositions were fitted for the stellar mass-to-light
ratio in the {\it R}-band varying from $0$
(minimum disk) up to $1.2$ M$_{\odot}$/L$_{\odot}^{R}$ (maximum disk).
We will denote by $\Upsilon_{\ast}$ the stellar mass-to-light ratio
in the {\it R}-band in solar units. It is noteworthy that 
$\Upsilon_{\ast}=1.2$ is the maximum value allowed by
the smooth rotation curve but avoiding a hollow halo in the inner two
points. However, it is submaximal
in the sense that the maximum contribution of the stellar disk to the
rotation curve is less than $0.8$ times the observed maximum rotation.
In practice, the stellar curve is scaled until the inner points match those
of the smooth curve. Depending on how many points are used, the maximum
disk mass-to-light ratios may vary as much as $\sim 15\%$.
In the case of NGC 5963 and using the inner five points, 
Simon et al.~(2004) obtained a dynamical value for the 
maximum mass-to-light ratio
larger than the value quoted above, but it appeared unrealistic
and inconsistent with the predictions
from the galaxy colours by the Bell et al.~(2003) population synthesis
models. The dynamical maximum disk $\Upsilon_{\ast}$ inferred by using
two points is fully consistent with these predictions.

\begin{figure}
\plotone{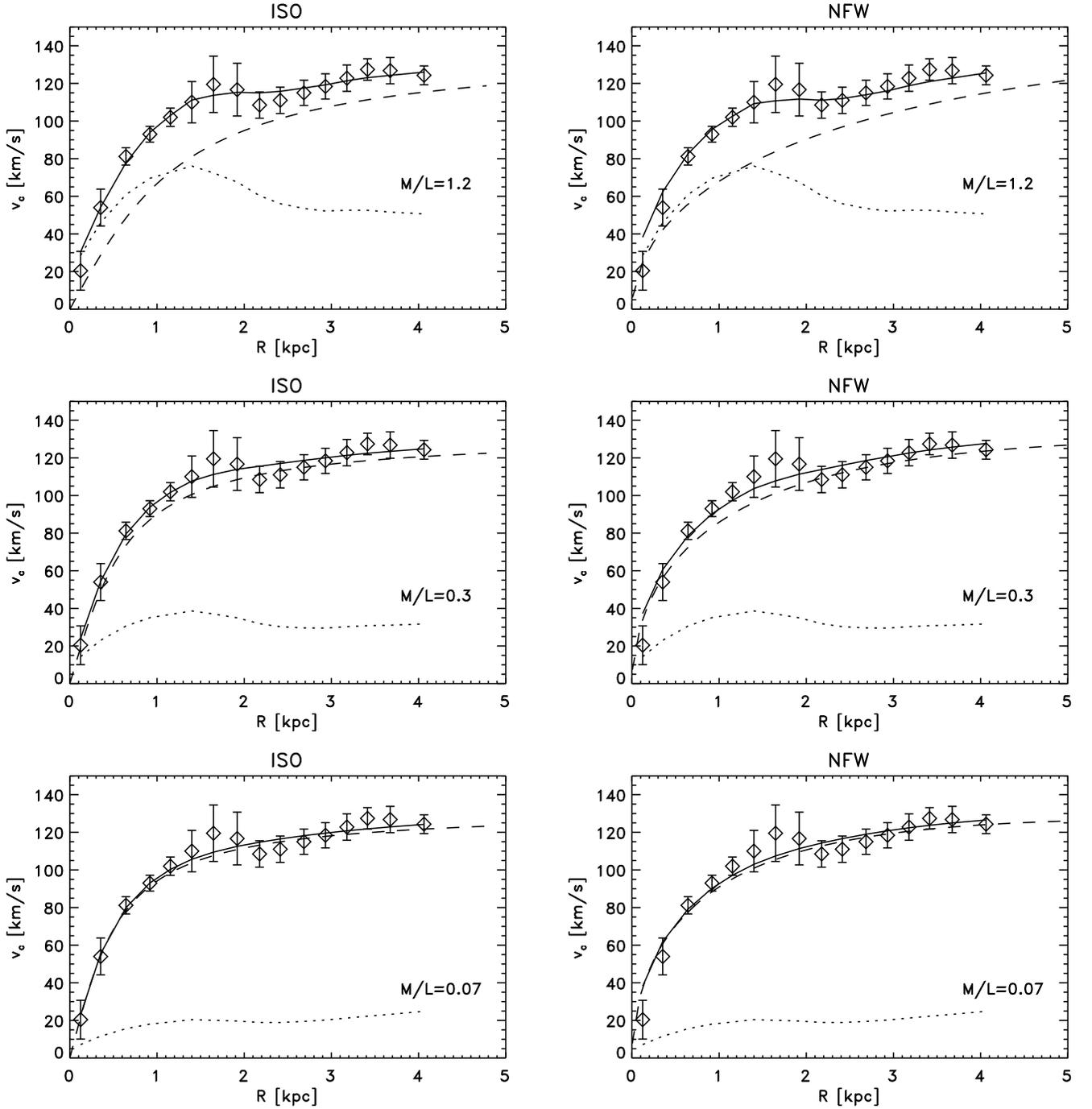}
\caption{Best-fitting mass models assuming pseudoisothermal halo (left) and
NFW halo (right), for different values of the stellar mass-to-light
ratio $M/L=\Upsilon_{\ast}$ in solar units. 
The contribution of the luminous matter (stars plus gas) in the disk
is shown as dotted lines. The resulting halo contribution (dashed line)
to the rotation curve and the final total model curve (solid line)
are also shown. The reduced $\chi^{2}$
of the fit and the probability $p$ that the data and the model could result
from the same parent distribution are given in Table \ref{table:parameters}. }
\label{fig:fits}
\end{figure}

Fig.~\ref{fig:fits} shows the best-fitting mass models 
for different values of
$\Upsilon_{\star}$ with both the NFW halo and the cored
pseudoisothermal halo: 
\begin{equation}
\rho'(r)=\frac{\rho_{0}' r_{\rm c}'^{2}}{r^{2}+r_{\rm c}'^{2}}.
\label{eq:pseudoiso}
\end{equation}
Primed quantities are used where appropriate to 
denote the (observed) final states after adiabatic compression
of the dark halo by the infall of the baryons as they cool and settle into
a disk. 
The parameters of each model, the reduced $\chi^2$ of the fit 
and the probability $p$ that the data
and the model could result from the same parent distribution are given
in Table \ref{table:parameters}.
It is important to remark that the observation that pseudoisothermal
fits produce low reduced $\chi^{2}$ values does not demonstrate the
presence of a constant-density core. Simon et al.~(2004) have shown that
a power-law density profile $\rho\propto r^{-1.2}$ fits also very well
the rotation curve of NGC 5963.

In order to compare with simulations which do not include the
baryonic physics, we need an
estimate of the DM profile before the adiabatic compression by
the baryon condensation, i.e.~the unprimed
parameters of the halo. A cored profile is expected after adiabatic
decompression.
Since the core baryons cause the dark profile
to steepen, we have fitted the observed rotation curve, for 
a given $\Upsilon_{\star}\geq 0.2$,
with a cuspy profile (for convenience we have chosen the NFW profile)
and recovered the halo parameters $(\rho_{0}, r_{c})$ of the approximately
pseudoisothermal sphere before compression,
following the standard model (e.g., Blumenthal et al.~1986;
Flores et al.~1993). For $\Upsilon_{\ast} < 0.2$, the effect of compression
by the baryons is so small that a cored profile is assumed before
and after contraction. 
The recipe of Blumenthal et al.~(1986) assumes that the matter distribution
is spherically symmetric and that particles move on circular orbits.
Jesseit et al.~(2002) verified that if the disk formation is smooth,
the contracted halos have density profiles that are in excellent
agreement with the standard predictions. If the disk formation involves
clumpy, cold streams or if the bulk of the central stars formed when
the halo mass was still being assembled, the level of halo contraction
may be smaller. Gnedin et al.~(2004) suggested an improvement of
the standard model of the response of the
halo to condensation of baryons 
by a simple modification of the assumed invariant (see Gnedin et al.~2004
for details). They show that the standard model systematically overpredicts
the contraction. Moreover, there exist other empirical evidence which
suggests that adiabatic contraction is avoided (e.g., Loeb \& Peebles 2003;
Dutton et al.~2005). Therefore, {\it our procedure
provides an upper limit on the effect of adiabatic contraction.}

\begin{figure}
\plotone{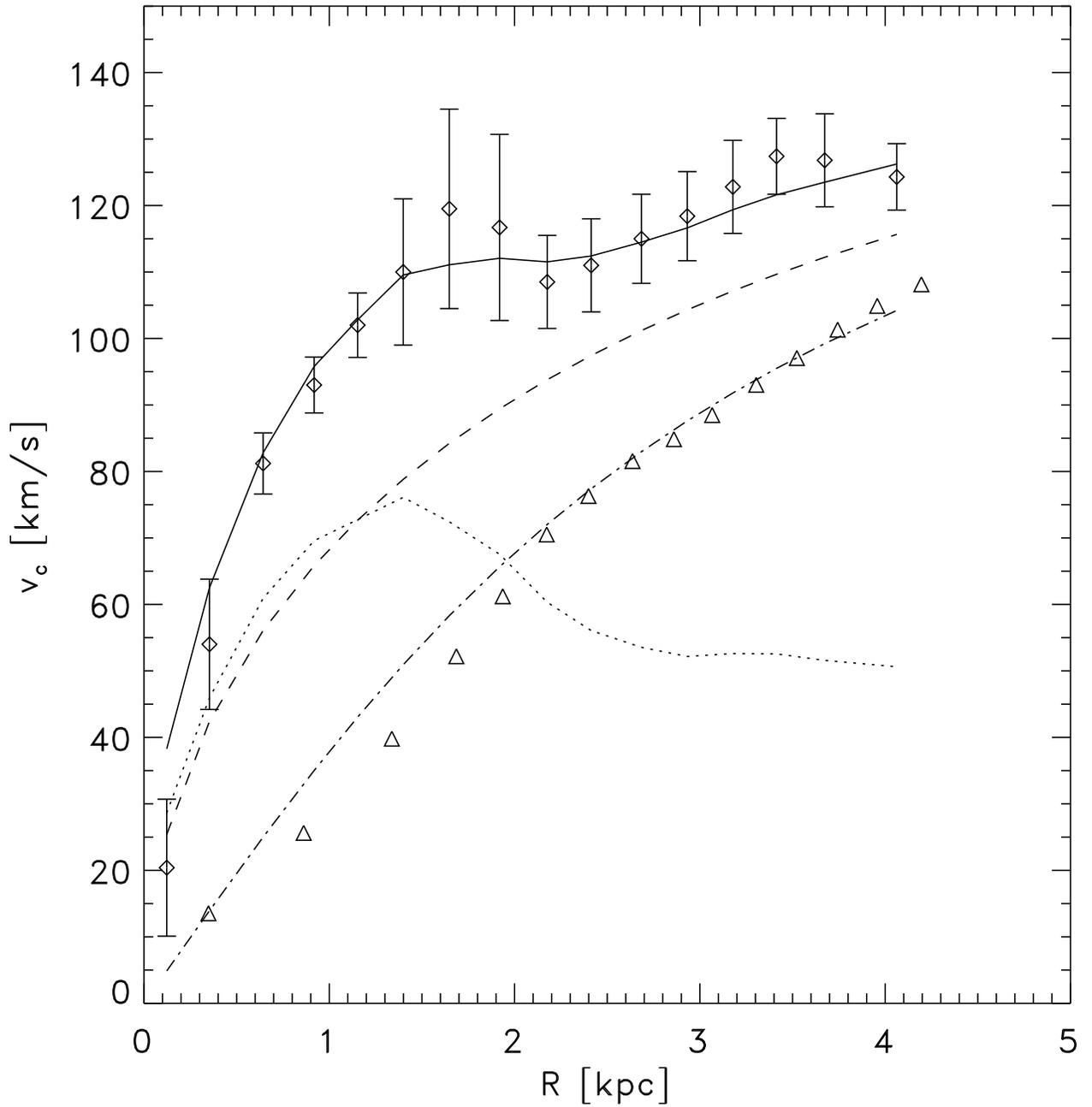}
\caption{Mass model with $\Upsilon_{\star}=1.2$
assuming NFW profile for the contracted halo. Symbols
as in Fig.~\ref{fig:fits}.
Dot-dashed line shows a fit to the dark matter rotation curve after
adiabatic decompression (triangles).
In this fit we adopted the pseudoisothermal profile.
The parameters of the halo before and after adiabatic contraction
are given in Table~\ref{table:parameters}. }
\label{fig:adbtc}
\end{figure}

For illustration, Fig.~\ref{fig:adbtc} shows the contribution to the
rotation curve of the dark halo after correcting by adiabatic contraction,
in the most extreme case of maximum disk $\Upsilon_{\star}=1.2$. 
The fit to it adopting the pseudoisothermal profile is also shown.
Note that if we demand that
the density of the halo must decrease monotonically with increasing
radius, values $\Upsilon_{\star}>1.2$ would not be permitted in the
{\it standard} adiabatic decompression approach.

In Table 1, best-fitting halo parameters are reported as
a function of $\Upsilon_{\star}$. Even adopting the maximum disk
solution, the halo of NGC 5963, at present time, 
is extraordinary concentrated, with a
central dark matter density greater than $0.35$ M$_{\odot}$ pc$^{-3}$
and a core radius $<1$ kpc. A thorough discussion aimed to identify
what about NGC 5963 makes it unique was given by Simon et al.~(2004)
and need not be repeated here. In the next sections,
we will discuss the implications of the halo of NGC 5963
for the self-interaction 
of DM particles.



\section{Constraints on the Cross-Section of DM Particles
at the Low Velocity Regime}
\label{sec:constraints}

In principle we do not really know whether the core of the galaxy 
under consideration is 
still in the expansion phase or undergoing gravothermal
core collapse aided by the gravitational contraction due to the baryons. 
In fact, the core contraction of the halo
by the collapsing baryons may trigger a
fast core collapse (e.g., Kochanek \& White 2000,
and \S \ref{sec:corecollapse}).
This effect may be important for $\Upsilon_{\ast}$ ratios close
to the maximum disk value.
Using cosmological simulations of SIDM as a calibrator,
we first put limits on the cross section of DM particles
assuming that the core of NGC 5963 is in the expanding phase
due to heat transfers inwards (\S \ref{sec:calibration}).
For large values of $\Upsilon_{\star}$, however, core collapse triggered by 
baryon condensation may have dramatic consequences for the longevity 
of the core. We have to demand that the halos are in
no danger of collapsing. This requirement implies another bound
on the cross section, which is discussed in \S \ref{sec:corecollapse}. 

The galaxy halo profiles and the
tight cross section limits coming from clusters of galaxies can
be reconciled only if the cross section were  
inversely proportional to the halo velocity
dispersion. Nevertheless, except for subhalos undergoing 
the action of the hot halo environment, 
most of the relevant results found in simulations 
with a cross section independent of halo velocity dispersion 
are valid if the value of the cross section is interpreted as that
appropriate to the velocity dispersion of the halo being considered.
Therefore, since we are primarily interested in a single object,
we can gain additional physical insight by making also use of simulations with
a velocity-independent cross section.



\subsection{Halos in Midsized Galaxies under SIDM: 
Calibrating Collisional Effects}
\label{sec:calibration}
In cosmological simulations of structure formation in SIDM scenario
with effective cross sections $0.5$--$1$ cm$^{2}$ g$^{-1}$,
midsized halos (masses $\sim 10^{10-11}$ M$_{\odot}$)
present cores with radii in the range $r_{c}\approx 2.5$--$5$ kpc and 
central densities $\rho_{0}\approx 0.01$--$0.06$ M$_{\odot}$
pc$^{-3}$ (Yoshida et al.~2000; Dav\'e et al.~2001;
Col\'{\i}n et al.~2002; D'Onghia et al.~2003). For instance, 
D'Onghia et al.~(2003) derive $r_{c}=5$--$6$ kpc
and $\rho_{0}\sim 0.02$ M$_{\odot}$ pc$^{-3}$ for a halo with
maximum circular velocity of 
$120$ km s$^{-1}$. Therefore, effective cross sections in the
range $0.5$--$1$ cm$^{2}$ g$^{-1}$ are suitable to reproduce soft
cores in late-type galaxies\footnote{In order to compare simulations, 
we notice that if the cross section
is assumed to be inversely proportional to the collision velocity, 
$\sigma_{\rm dm}=\sigma_{\ast}(v_{0}/v_{\rm rel})$,
the effective cross section for a halo with (one-dimensional) velocity 
dispersion $\tilde{v}$ is $\sigma_{\ast}/(\sqrt{\pi}\tilde{v}_{100})$,
where $\tilde{v}_{100}=(\tilde{v}/100\,{\rm km \,s}^{-1})$. To show
this relation one has to remind that, for a Maxwellian velocity distribution,
$\left<1/v_{\rm rel}\right>=1/(\sqrt{\pi}\tilde{v})$.}. 

All the simulations with $\sigma_{\rm dm}=\sigma_{\ast}(v_{0}/\tilde{v})$
also shown that for $\sigma_{\ast}=0.5$--$1$ cm$^{2}$ g$^{-1}$,
the number of collisions per particle
per Hubble time at the center of the halos, was between $4$--$6$,
roughly independent of the halo mass.
In fact, while a few collisions (between $2$ and $3$) at the halo 
center are enough
to produce a constant density core (Yoshida et al.~2000), 
$4$--$6$ collisions are sufficient to produce a core with 
$r_{c}\sim 0.4 r_{s0}$,
with $r_{s0}$ the scale radius of the NFW halo when
it is resimulated in the standard, collisionless 
case ($\sigma_{\rm dm}=0$). The reason is that each scattering
produces a change $\Delta v\sim v$, and particles escape the core
in a single scattering.

Rescaling the number of collisions per particle at the center
of the cluster labelled S1Wa in Yoshida et al.~(2000), we find 
\begin{equation}
N_{\rm col}\approx 3 \left(\frac{\rho_{0}}
{0.02 \,{\rm M}_{\odot} {\rm pc}^{-3}}\right)
\left(\frac{\tilde{v}}{100\, {\rm km}\, {\rm s}^{-1}}\right)
\left(\frac{\sigma_{\rm dm}}{1\, {\rm cm}^{2} {\rm g}^{-1}}\right).
\label{eq:collisions}
\end{equation}
We stress here that $\rho_{0}$ is the final unprimed central density of the
core, after thermalization. Let us estimate $N_{\rm col}$ in the
halo center of NGC 5963 for mass models with $\Upsilon_{\ast}<0.7$.
For a DM cross section
$\sigma_{\rm dm}\tilde{v}_{100}$ between $0.5$ and $1.0$ cm$^{2}$ g$^{-1}$, 
where $\tilde{v}_{100}=(\tilde{v}/100\, {\rm km\,s}^{-1})$, and
a central density $\rho_{0}\approx 0.4$ M$_{\odot}$ pc$^{-3}$,
the mean number of collisions is 
$N_{\rm col}\approx 20$--$45$.
Hence, we expect NGC 5963 having a core radius greater than $0.4 r_{s0}$.
In fact, from the scaling laws of Yoshida et al.~(2000), we see
that the core radius reaches a size $\sim 1.5 r_{s0}$ when the collision
rate per particle at halo center is $\sim 30$ per Hubble time.
In order to quantify the expected size of the core of this galaxy
under SIDM, we need its characteristic radius $r_{s0}$.

An estimate of $r_{s0}$ can be inferred for NGC 5963
by fitting the rotation curve beyond $5$ kpc with a NFW profile.
The observed inner rotation curve cannot be included as it is altered 
as a consequence of DM self-interactions and the adiabatic contraction
that causes the DM profile to steepen. Since there are many combinations
of the parameters $(c, V_{200})$ that can fit the rotation curve,
we need to choose either $c$ or $V_{200}$. Expected values of
the concentration in $\Lambda$CDM cosmology have a $2\sigma$ range
from $5$ to $22$ (Eke et al.~2001; Jing \& Suto 2002). 
We will take a value $c=20$, which lies among the largest values
in the $2\sigma$ uncertainty. The exact value of $r_{s0}$ also 
depends slightly on $\Upsilon_{\star}$. For the intermediate case 
$\Upsilon_{\star}=0.6$ (and $c=20$), we find that 
the characteristic radius is $r_{s0}\approx 6$ kpc.
For the average value predicted by simulations, $c=13$, 
we found $r_{s0}=11.7$ kpc. Hence, 
as a conservative calibration, we will assume
that for $N_{\rm col}\gtrsim 6$, a core radius $r_{c}\approx
0.4 r_{s0}\geq 2.5$ kpc should have been formed. 

In Table \ref{table:parameters} we see that for models with 
$\Upsilon_{\ast}<0.7$, the inferred core radius before
contraction in NGC 5963 is smaller than $1.2$ kpc; the question that arises
is how many collisions suffice to develop a core with $r_{c}\lesssim 1$ kpc. 
To answer this question we have examined the SIDM simulations 
of the evolution of a Hernquist halo with total mass $M_{T}$ 
and break radius $r_{H}$ in Kochanek \& White (2000).  
These authors carried out simulations 
surveying the dimensionless cross section
$\hat{\sigma}_{\rm dm}=M_{T}\sigma_{\rm dm}/r_{H}^2$. 
According to their Fig.~2c, the core density
in the phase of core expansion, and for halos with different
$\hat{\sigma}_{\rm dm}$, goes as 
$\propto \hat{\sigma}_{\rm dm}^{-\eta}$, with $\eta <1/2$.
Moreover, at the time that the core of the simulation 
with dimensionless cross section
$\hat{\sigma}_{\rm dm}=3$ reaches its maximum radius, the run
$\hat{\sigma}_{\rm dm}=0.3$ has already developed a core  
of half this radius,
with a mean collision count at halo center $\sim 5$ times smaller.
Applying this scaling to NGC 5963 it holds that 
a mean of $6/5$ collisions per particle at the center,
will suffice to produce a core of $2.5/2$ kpc.
Although these estimates are based on simulations of the relaxation
of an isolated galaxy having initially a cuspy Hernquist profile, 
it was also found in cosmological simulations that
a few collisions (exceeding $2$) suffice to develop a kpc-size core
(e.g., Yoshida et al.~2000). 
To keeps matter simple we will adopt the {\it generous condition} that only
if $N_{\rm col}\leq 2$ then $r_{c}\lesssim 1$ kpc. 



Putting together, 
we take the following, rather conservative, relationships
as a calibrator of the collisional effects:
\begin{equation}
{\rm if}\hskip .14cm N_{\rm col}= 6 \Leftrightarrow \hskip .1cm r_{c}=2.5 
\hskip .1cm
{\rm kpc},
\label{eq:calib1}
\end{equation} 
\begin{equation}
{\rm if}\hskip .14cm N_{\rm col}= 2 \Leftrightarrow \hskip .1cm r_{c}\leq 1.0 
\hskip .1cm
{\rm kpc}.
\label{eq:calib2}
\end{equation} 
Suppose that $\sigma_{\rm dm}\propto \tilde{v}^{-1}$. The generalization
for a power-law $\sigma_{\rm dm}\propto \tilde{v}^{-a}$ is obvious and
it will be ignored in the interest of simplicity.
To place an upper limit on $\sigma_{\rm dm}\tilde{v}_{100}$, we proceed
as follows. We first derive the best-value fitting parameters 
($r_{c},\rho_{0})$ for a given $\Upsilon_{\star}$. 
If $1$ kpc $<r_{c}<3$ kpc, we estimate $N_{\rm col}$
by interpolating Eqs.~(\ref{eq:calib1})-(\ref{eq:calib2}) linearly.
Once $N_{\rm col}$ is known,
Equation (\ref{eq:collisions}) immediately provides us with an upper limit on
$\sigma_{0}\equiv\sigma_{\rm dm} \tilde{v}_{100}$.
If $r_{c}$ is smaller than $1$ kpc,
we will assume that $N_{\rm col}\leq 2$,
which gives:
\begin{equation}
\sigma_{0}\leq 0.7 {\rm cm}^{2}{\rm g}^{-1}
\left(\frac{\rho_{0}}{0.02 {\rm M}_{\odot} {\rm pc}^{-3}}
\right)^{-1}.
\end{equation}
For instance, if we have a situation in which $r_{c}<1$ kpc and 
$\rho_{0}=0.4$ M$_{\odot}$ pc$^{-3}$, 
we would obtain $\sigma_{0}\leq 0.035$ cm$^{2}$ g$^{-1}$.


\subsection{Results}
\begin{figure}
\plotone{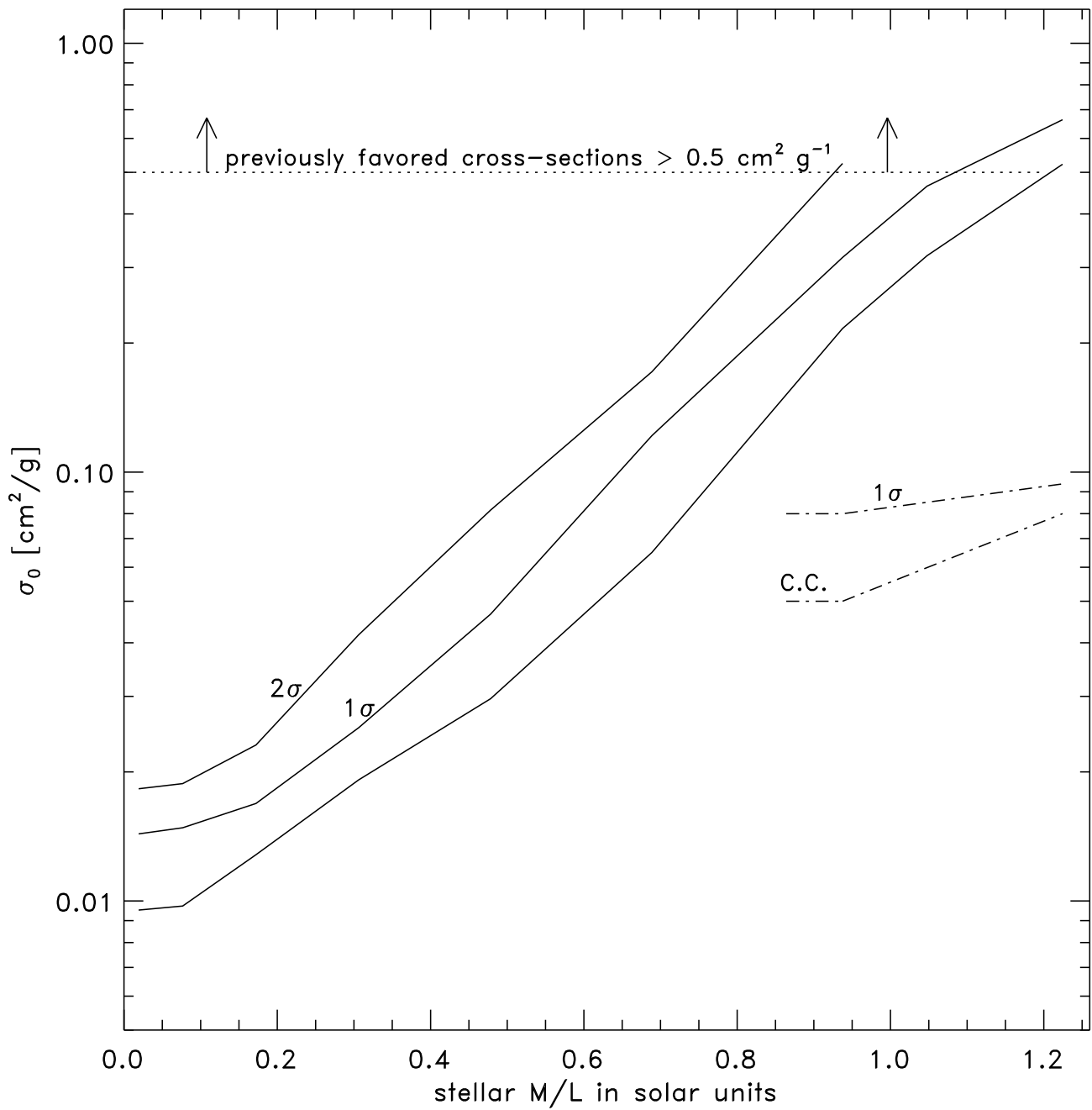}
\caption{Exclusion contour plot for the dark matter interaction
as a function of $\Upsilon_{\ast}$.
Various confidence levels are shown. Solid curves trace the upper limit
on $\sigma_{0}$ derived by the constraint discussed in \S 
\ref{sec:calibration}.
Dot-dashed lines are the core collapse (C.C.) constraint, obtained by
requiring that the lifetime of the halo of NGC 5963 is longer than
one Hubble time (\S \ref{sec:corecollapse}). }
\label{fig:results}
\end{figure}

Figure \ref{fig:results} shows the resulting upper limits on $\sigma_{0}$
versus $\Upsilon_{\star}$, 
following the procedure described in the
previous subsection. Confidence levels were calculated as a measure
of the sensitivity of the results to the goodness of the fit to the
rotation curve. Other sources of uncertainties in the
parameters of the model, as the distance to the galaxy,
its inclination, or the adopted characteristic 
radius $r_{s0}$, were not taken into account in these contours. 
We must warn that these error bands cannot be considered as real probability
indicators because the velocities and their errors are not free
of systematic effects. 

Our derivation overestimates the upper
limit on $\sigma_{0}$ 
because we have assumed in the derivation of the halo parameters that
baryons only act to hasten core contraction. Nevertheless, other
phenomena as galactic bars, outflows, massive black holes  
in the halos, and other processes associated
with AGN or star formation have been proposed as mechanisms to
erase cuspy halos. We are being conservative in ignoring these phenomena
because incorporating their effects would only serve to place more
stringent constraints on $\sigma_{0}$.

The bound on $\sigma_{0}$ depends strongly on $\Upsilon_{\ast}$.
From Fig.~\ref{fig:results} we see that models with
$\Upsilon_{\ast} \approx 0.6$ are consistent with $\sigma_{0}$ just under
$0.1$ cm$^{2}$ g$^{-1}$. For $\Upsilon_{\star}=0.35$, 
$\sigma_{0}>0.05$ cm$^{2}$ g$^{-1}$ can be excluded virtually at 
$2\sigma$ confidence interval.  
Within the range $0<\Upsilon_{\ast}<0.7$, our limit excludes
the interval proposed to explain the flat mass profiles in galaxies.

For $\Upsilon_{\star}>1.0$, the present analysis only disfavours the interval
of astrophsical interest, $\sigma_{0}=0.3$--$1$ cm$^{2}$ g$^{-1}$, 
at less than $68\%$ confidence level. 
We should notice, however, that if the contribution of the
baryons to the potential well is important, the core may be in danger
of undergoing gravothermal collapse. 
The requirement that the core is not undergoing a fast collapse will
provide a more stringent upper limit for the DM cross section
for $\Upsilon_{\star}>0.8$.

\subsection{Core Collapse}
\label{sec:corecollapse}
In the latter section we found that a cross section 
$\sigma_{0}\sim 0.5$ cm$^{2}$ g$^{-1}$ and $\Upsilon_{\star}>1.0$
would be marginally consistent with a picture in which the 
halo of NGC 5963 might have formed a core with a radius 
$\sim 2$--$2.5$ kpc, although significantly reduced because of the
adiabatic contraction by baryons. Nevertheless, we must also require that this
galaxy has yet to undergo core collapse. 
For these high $\Upsilon_{\ast}$, 
the cooling baryons will compress the dark matter and accelerate
the core collapse in different ways (e.g., Kochanek \& White 2000).
By raising the density due to the contraction, the relaxation time
drops. In addition, the adiabatic compression produces a steeper
central density cusp which further speeds up the ultimate evolution
of the system. Moreover, the initial inversion of the temperature
(velocity dispersion of DM particles) profile responsible for the
core expansion may be erased by the adiabatic heating of DM particles,
leading directly to a core collapse with no room for the expansion phase. 
We need to estimate the characteristic time for core collapse  
in the case $\Upsilon_{\star}\geq 0.8$, i.e.~when the effects of 
contraction are important. In their numerical simulations, 
Kochanek \& White (2000) assumed as initial conditions a Hernquist 
profile with a cuspy core. In our case, i.e.~for values
$\Upsilon_{\star}\geq 0.8$, it is realistic to
start with such an initial configuration because a cusp in the
density profile is a natural consequence of the adiabatic
contraction caused by the baryons.
In this regard, we can make use of the quantitative study of 
Kochanek \& White (2000).   

The evolution of the collapse depends on the ratio between the
collision mean free path, $\lambda$, to the local gravitational
scale height $H$ (Balberg et al.~2002). Systems initially having 
ratios $(\lambda/H)_{\rm edge}\lesssim 10$ at the outer edge of the core,  
have a lifetime of only a few relaxation times\footnote{It is 
simple to show that
$\hat{\sigma}_{\rm dm}$ and $(\lambda/H)_{\rm edge}$ are related by the
relation $(\lambda/H)_{\rm edge}\approx 2.2/\hat{\sigma}_{\rm dm}$.} 
(Quinlan 1996; Kochanek \& White 2000; Balberg et al.~2002).
For the range of $\hat{\sigma}_{\rm dm}$ explored in Kochanek \&
White (2000), the central density increases by
an order of magnitude in a timescale 
$t_{c,10}\sim 4 \hat{\sigma}_{\rm dm}^{1/2}t_{\rm rc}$,
being $\hat{\sigma}_{\rm dm}=2\pi \rho_{i}r_{s}\sigma_{\rm dm}$
for the NFW profile.
In particular, for $\hat{\sigma}_{\rm dm}=0.3$, this timescale is
only $\sim 2.2$ times the core-radius relaxation time.
This result is troublesome for the survival of the core because
it recollapses quickly after its formation and could not
persist until today (see \S \ref{sec:basic}).

Let us calculate the core timescale $t_{c,10}$ for the halo of NGC 5963.
As said before, because of the adiabatic compression by the baryons,
a cuspy halo, such as the NFW profile, is expected even in the presence of
DM collisions. The best NFW model for $\Upsilon_{\star}=1.2$ 
corresponds to $c=15.6$ and $V_{200}=109$ km s$^{-1}$, implying
a core-radius timescale:   
\begin{equation}
t_{\rm rc}=\frac{1}{3\rho_{i}\sigma_{\rm dm}\tilde{v}}=0.8\,{\rm Gyr}
\left(\frac{\sigma_{0}}{1\,{\rm cm}^{2}{\rm g}^{-1}}\right)^{-1}.
\end{equation}
For the above halo parameters, 
$\hat{\sigma}_{\rm dm}=2\pi \rho_{i}r_{s}\sigma_{\rm dm}=
0.15\,\tilde{v}^{-1}_{100}(\sigma_{0}/1\,{\rm cm}^{2}\,{\rm g}^{-1})$ and
hence the central density will be enhanced by a factor $10$ in 
the timescale
\begin{equation}
t_{c,10}\simeq 4\hat{\sigma}_{\rm dm}^{1/2}t_{\rm rc}=
1.25\,{\rm Gyr} \,\left(\frac{\tilde{v}}{100\,{\rm km\, s}^{-1}}\right)^{-1/2}
\left(\frac{\sigma_{0}}{1\,{\rm cm}^{2}{\rm g}^{-1}}\right)^{-1/2}.
\end{equation}

Taken $\sigma_{0}=0.5$ cm$^{2}$ g$^{-1}$, we get 
$t_{c,10}\approx 2$ Gyr. This estimate has uncertainties of as much
as a factor $2$. 
This short timescale indicates that the dark halo may have sufficient time to
increase its central density by a factor $100$-$1000$.
In order to prevent the dark halo from such dramatic evolution,
we must demand that $t_{\rm rc}\sim$ one Hubble time, 
implying $\sigma_{0}\lesssim 0.08$
cm$^{2}$ g$^{-1}$. Proceeding in the same way 
we obtain $\sigma_{0}\lesssim 0.05$ cm$^{2}$ g$^{-1}$
but now for the mass model with $\Upsilon_{\star}=1.0$.
The line connecting these two points in Fig.~\ref{fig:results} 
delimits the region
where cross sections are probably small enough to avoid catastrophic
core collapse in the halo of NGC 5963. The corresponding line 
at the $1\sigma$ confidence is also plotted.
We see that the new constraint, which accounts for the role of baryons
when $\Upsilon_{\ast} >0.85$, is tighter than the one derived in the
previous subsections. This all but removes the permitted window
for cross sections $\sigma_{0}=0.3$--$1$ cm$^{2}$ g$^{-1}$ in mass
models with $\Upsilon_{\ast}>0.85$, that remained open.
Combining both constraints, the maximum of the permitted value
$\sigma_{0}\approx 0.2$ cm$^{2}$ s$^{-1}$, at $2\sigma$ confidence level, 
occurs for $\Upsilon_{\ast}\approx 0.7$. 

\section{Discussion and conclusions}
\label{sec:discussion}
Apart from the obvious interest for the still unknown nature of
dark matter, the possibility of it having a nonzero self-interaction
cross section has other astrophysical implications.  
SIDM was suggested as a route to form cores in LSB and dwarf galaxies.
The fact that the required cross sections are comparable to the cross section
for particles interacting with each other via the strong force, has
led to speculate that DM particles could interact with both
themselves and with baryons through the strong force (Wandelt et al.~2001). 
The interaction of dark matter with protons might contribute
to reheat the intracluster medium in the central regions
of clusters of galaxies (Qin \& Wu 2001; 
Chuzhoy \& Nusser 2004).

A large range of parameters space is being ruled out by
current experimental and astrophysical bounds.
Previous studies have gradually whittled down the DM cross section
allowed to solve the cuspy problem of halos in $\Lambda$CDM cosmology.
In addition to statistical studies, the analysis of individual galaxies 
can give additional constraints on the strength of DM self-interaction
cross section.

The dark matter distribution in NGC 5963 is challenging
for any model designed to form central cores in dwarf
and LSB galaxies. Apparently, its halo distribution  
seems at odds with the results of SIDM 
simulations with cross sections $\sigma_{\rm dm}\tilde{v}_{100}=0.3$--$1$ 
cm$^{2}$ g$^{-1}$,
which produce central densities $\sim 0.02$ M$_{\odot}$ pc$^{-3}$,
fairly independent of the halo mass, and core radii $2.5$--$5$ kpc.
The highly concentrated halo of NGC 5963 
implies upper limits for the interaction
of dark matter particles, unless we change our assumption of constant
$\Upsilon_{\ast}$. A higher $\Upsilon_{\ast}$ for the disk than
for the lens is required to have a less concentrated halo. However, 
the observed {\it B-V} colours of the disk
and the lens do not favour this possibility (see Bosma et al.~1988 for
a discussion). 

Great efforts have been made to constrain the collisional cross section
of DM particles. It was suggested that this cross section should be
small enough so the core halo of dwarf and LSB galaxies would not
collapse in a Hubble time (e.g., Hennawi \& Ostriker 2002). Hence,
the halo of NGC 5963 by itself cannot be going through the gravothermal 
contraction phase. However, core collapse may be induced by the action
of the baryons as they deepen the potential well and compress the 
core in an adiabatic process. Therefore,
the core of NGC 5963 may be expanding if the thermalization
has not been completed, or shrinking if the core is adiabatically compressed,
which may occur only at large values of $\Upsilon_{\ast}$.


Either the core of NGC 5963 is expanding or contracting in size, 
a tight constraint
$\sigma_{0}<0.2$ cm$^{2}$ g$^{-1}$ at $95\%$ confidence level, 
which indicates that our results are robust to reasonable DM uncertainties,
is derived.
This upper limit for $\sigma_{0}$ may be overestimated 
as much as a factor $2$ because of our conservative assumptions. 
Thus, the original motivation for SIDM of lowering the core densities
of galactic halos require collisional cross sections too large to be
consistent with the halo of NGC 5963.

One of the largest uncertainties in any mass model is the precise
value of $\Upsilon_{\ast}$ because depends on extinction, star
formation history, etc. Based on various considerations, e.g., stellar
population synthesis, stellar counts and kinematics in the
solar neighbourhood and kinematics of external galaxies,
some authors argue that $\Upsilon_{\ast}^{I}\approx 2h$ in the {\it I}-band 
(e.g., Mo \& Mao 2000, and references therein) and
$\Upsilon_{\ast}^{R}\approx 1.2$ in the {\it R}-band (Simon et al.~2004). 
This would mean that the
constraint derived from the core collapse discussed in \S \ref{sec:corecollapse}
is the most relevant, and suggests an upper limit $\sigma_{0}\lesssim 0.1$
cm$^{2}$ g$^{-1}$. This value coincides with the upper limit inferred
in clusters of galaxies (e.g., Meneghetti et al.~2001; Arabadjis \&
Bautz 2004).

Upper limits in the $0.02$--$0.1$ cm$^{2}$ g$^{-1}$ range were derived by
Hennawi \& Ostriker (2002) from the mass of supermassive black
holes in the centers of galaxies.
For effective cross sections $\sigma_{\rm dm}>0.02$ 
cm$^{2}$ g$^{-1}$, the accretion of
SIDM onto seed black holes would produce excessively massive black holes.
However, there are serious uncertainties
associated with this limit because they make use of the
hypothesis that initially the density DM distribution follows
the NFW profile, whereas 
the numerical study of Col\'{\i}n et al.~(2002) of SIDM cosmology
suggests that the NFW profile at the center of the halos
is not achieved at any time. Therefore,
the otherwise excessive accretion of matter onto the black hole
 may not occur (Col\'{\i}n et al.~2002). 

In order to reconcile the collisional DM hypothesis as a viable explanation
of the formation of the constant density cores with the 
halo of NGC 5963, we should identify potential ways able to compensate the
evacuation of DM in the central parts caused by the scattering of DM or
to forestall the core collapse, depending whether the core of
NGC 5963 is expanding or collapsing. 
Tidal redistribution of mass at the halo center cannot be efficient
in NGC 5963 because the nearest large galaxy in the group is at a
projected distance of $430$ kpc. One could relax our simplifying
assumption that the halo consists of a well mixed homogeneous DM
distribution and to consider a clumpy medium. 
The final density profile will be the result
of two competing effects. 
On the one hand, the mass infall associated
with the spiraling of putative massive clumps of DM 
($\gtrsim 10^{6-7}$ M$_{\odot}$) towards the galactic center by
dynamical friction, produces a
replenishment of material and deepens the central potential well. On the
other hand, dynamical friction heating may be effective in flattening
the inner DM profile. Unlike clusters of galaxies in which 
substructure is important in determining the final DM distribution
(e.g., Nipoti et al.~2004), halos of LSB galaxies are already largely
assembled at $z\approx 3$ and hence, they have had sufficient time to 
disrupt the halo substructure, forming a smooth DM distribution within the
luminous radius, unless NGC 5963 had a prominent mass aggregation
history. However, an anomalous accretion history would be very
unsatisfactory because it would have dramatic consequences for 
the disk itself, producing an excessive dynamical
heating or even its destruction. 
Moreover, Ma \& Boylan-Kolchin (2004) argued that energy deposition
by merging dark matter substructures likely flatten density profiles. 
In the lack of any of those potential mechanisms capable to make
significant mass redistribution, we think that
our analysis is robust, even though it is based on the halo
of a single galaxy.

The halo of NGC 5963 is problematic for any model whose mechanism to
produce large cores in LSB galaxies
depends on collisions between DM particles (e.g., annihilating dark matter),
reducing the parameters space and suggesting new directions for DM search.
Our analysis also places strong constraints to the non-hadronic exotic
Q-balls as a dark matter candidate in galaxies. These ``particles''
might be arranged to have mutual collisions with a large cross section and,
in addition, Q-balls can stick together after
collision, reducing the self-interaction as scattering proceeds
(Kusenko \& Steinhardt 2001). In order
to be consistent with the halo of NGC 5963, self-interactions between
Q-balls should shut off to a negligible value after $1$--$2$ collisions
per particle in order for the initial scatterings to smooth out
halo cusps but avoiding gravothermal collapse.

Perhaps the solution of the halo core problem resides in decaying
dark matter (Cen 2001a,b; S\'anchez-Salcedo 2003). If 
dark matter particles in galactic halos
decay to stable particles with a recoiling velocity of a few tens
of kilometers per second, then a fraction of LSB galaxies can still present
a substantial concentration. This novel dynamics associated
with decaying dark matter was illustrated by S\'anchez-Salcedo (2003) 
for the case of the LSB galaxy NGC 3274.
We believe that the halos of NGC 5963 and NGC 3274 are not pathological
cases as far as their DM distributions concern. 
There are additional independent evidence on the existence of
galaxies with large central densities. Loewenstein \& Mushotzky
(2002), in an as-yet-unpublished work, have determined the 
enclosed mass profile for the elliptical galaxy NGC 4636 using X-ray 
observations. For this galaxy the central
density in models with dark matter cores is higher than
expected in the SIDM scenario with $a\leq 1$ (Loewenstein \& Mushotzky
2002). 

Future observations of LSB galaxies containing a very low density
of luminous material even in the inner parts, so that the observed
dynamics should be dominated by the gravitational forces of the dark
halo at small radii as well as large radii, will be able
to determine whether cores are produced by the gravitational interaction
between the luminous and dark matter, and will provide further constraints
on the nature of DM in galaxies.

\acknowledgements
I thank Tony Garc\'{\i}a Barreto and Alberto Bolatto for helpful discussions.
I am grateful to the anonymous referee for constructive comments.
This work was supported by CONACYT project 2002-C40366.

\clearpage
\begin{table*}
\caption[]{Fitting parameters for pseudoisothermal (ISO) and NFW halos
after and before halo contraction.}
\vspace{0.01cm}

\begin{center}
\begin{tabular}{c c c c c |c c}
\noalign{\medskip\hrule\medskip}
 & & NFW & & & \multicolumn{2}{c}{ISO after decompression}\\
\hline\hline
  { $\Upsilon_{\ast}$} & {$c'$} & 
{$V_{200}'$} & {$\chi^{2}_{\rm red}$} &
{$p$} &  { $r_{c}$} & {$\rho_{0}$} \\
  &  & (km s$^{-1}$) & & & {(kpc)}& {(M$_{\odot}$ pc$^{-3}$)}\\

\hline
    1.2   & 15.6  & 109.0 &  0.50  & 0.934  &  2.60   & 0.080 \\
    0.7   & 23.2  & 88.0 &  0.44  & 0.962  & 1.20   & 0.260 \\
    0.3   & 28.0  & 82.0 &  0.60  & 0.867  & 0.71   & 0.696 \\
    0.07  & 33.0  & 76.0 &  0.63  & 0.842  &  --   &  -- \\
\noalign{\medskip\hrule\medskip}
& & ISO & & & & \\
\hline\hline
 { $\Upsilon_{\ast}$} & {$r_{c}'$} &
{$\rho_{0}'$} & {$\chi^{2}_{\rm red}$} &
{$p$} & \\
 & (kpc) & (M$_{\odot}$ pc$^{-3}$) & & & \\
\hline
  1.2   & 0.95  & 0.38 &  0.29  & 0.995  & --  & -- \\
  0.3   & 0.48  & 1.35 &  0.28  & 0.996  & --  & -- \\
  0.07  & 0.42  & 1.75 &  0.28  & 0.996  & 0.49  &  1.375 \\
\hline
\end{tabular}
\end{center}
\label{table:parameters}
\end{table*}

\clearpage

\clearpage

\clearpage
\clearpage

\begin{thebibliography}{}
\bibitem[Arabadjis \& Bautz(2004)]{ara04}
Arabadjis, J. S., \& Bautz, M. W. 2004, astro-ph/0408362
\bibitem[Arabadjis, Bautz \& Garmire(2002)]{ara02}
Arabadjis, J. S., Bautz, M. W., \& Garmire, G. P. 2002, \apj, 572, 66
\bibitem[Balberg, Shapiro \& Inagaki(2002)]{bal02}
Balberg, S., Shapiro, S. L., \& Inagaki, S. 2002, \apj, 568, 475
\bibitem[Bell et al.(2003)]{bel03}
Bell, E. F., McIntosh, D. H., Katz, N., \& Weinberg, M. D. 2003, \apjs,
149, 289
\bibitem[Blumenthal et al.(1986)]{blu86}
Blumenthal, G. R., Faber, S. M., Flores, R., \& Primack, J. R. 1986,
\apj, 301, 27
\bibitem[Bolatto et al.(2003)]{bol03}
Bolatto, A. D., Simon, J. D., Leroy, A., \& Blitz, L. 
2003 in Dark Matter in Galaxies, IAU Symp.~220,
eds.~S. D. Ryder, D. J. Pisano, M. A. Walker, \& K. C. Freeman, 
San Francisco: Astronomical Society of the Pacific, 353
\bibitem[Bosma, van der Hulst \& Athanassoula(1988)]{bos88}
Bosma, A., van der Hulst, J. M, \& Athanassoula, E. 1988, \aap, 198, 100
\bibitem[Burkert(2000)]{bur00}
Burkert, A. 2000, \apj, 534, L143
\bibitem[Cen(2001a)]{cen01a}
Cen, R. 2001a, \apj, 546, L77
\bibitem[Cen(2001b)]{cen01b}
Cen, R. 2001b, \apj, 549, L195
\bibitem[Chuzhoy \& Nusser(2004)]{chu04}
Chuzhoy, L, \& Nusser, A. 2004, astro-ph/0408184
\bibitem[Col\'{\i}n, Avila-Reese, Valenzuela \& Firmani(2002)]{col02}
Col\'{\i}n, P., Avila-Reese, V., Valenzuela, O., \& Firmani, C.
2002, \apj, 581, 777 
\bibitem[Dav\'e et al.(2001)]{dav01}
Dav\'e, R., Spergel, D. N., Steinhardt, P. J., \& Wandelt B. D. 2001,
\apj, 547, 574
\bibitem[de Blok \& Bosma(2002)]{deb02}
de Blok, W. J. G., \& Bosma, A. 2002, \aap, 385, 816
\bibitem[D'Onghia, Firmani \& Chincarini(2003)]{don03}
D'Onghia, E., Firmani, C., \& Chincarini, G. 2003, \mnras, 338, 156
\bibitem[Dutton et al.(2005)]{dut05}
Dutton, A. A., van den Bosch, F. C., Courteau, S., \& Dekel, A.
2005, astro-ph/0501256, in Baryons in Dark Matter Halos, eds.~R.-J.~Dettmar,
U.~Klein, P.~Salucci
\bibitem[Eke, Navarro \& Steinmetz(2001)]{eke01}
Eke, V. R., Navarro, J. F., \& Steinmetz, M. 2001, \apj, 554, 114
\bibitem[Firmani et al.(2000)]{fir00}
Firmani, C., D'Onghia, E., Avila-Reese, V., Chincarini, G., \&
Hern\'andez, X. 2000, \mnras, 315, L29
\bibitem[Flores et al.(1993)]{flo93}
Flores, R., Primack, J. R., Blumenthal, G. R., Faber, S. M.
1993, \apj, 412, 443 
\bibitem[Gnedin et al.(2004)]{gne04}
Gnedin, O. Y., Kravtsov, A. V., Klypin, A. A., \& Nagai, D. 2004,
\apj, 616, 16
\bibitem[Gnedin \& Ostriker(2001)]{gne01}
Gnedin, O. Y., \& Ostriker, J. P. 2001, \apj, 561, 61
\bibitem[Hennawi \& Ostriker(2002)]{hen02}
Hennawi, J. F., \& Ostriker, J. P. 2002, \apj, 572, 41
\bibitem[Hernquist(1990)]{her90}
Hernquist, L. 1990, \apj, 356, 359 
\bibitem[Jesseit, Naab \& Burkert(2002)]{jes02}
Jesseit, R., Naab, T., \& Burkert, A. 2002, \apj, 571, L89
\bibitem[Jing \& Suto(2000)]{jin00}
Jing, Y. P., \& Suto, Y. 2000, \apj, 529, L69
\bibitem[Kochanek \& White(2000)]{koc00}
Kochanek, C. S., \& White, M. 2000, \apj, 543, 514
\bibitem[Kusenko \& Steinhardt(2001)]{kus01}
Kusenko, A., \& Steinhardt, P. J. 2001, \prl, 87, 141301
\bibitem[Lewis, Buote \& Stocke(2003)]{lew03}
Lewis, A. D., Buote, D. A., \& Stocke, J. T. 2003, \apj, 586, 135
\bibitem[Loeb \& Peebles(2003)]{loe03}
Loeb, A., \& Peebles, P. J. E. 2003, \apj, 589, 29
\bibitem[Loewenstein \& Mushotzky(2002)]{loe02}
Loewenstein, M., \& Mushotzky, R. 2002, astro-ph/0208090
\bibitem[Ma \& Boylan-Kolchin(2004)]{ma04}
Ma, C., \& Boylan-Kolchin, M. 2004, \prl, 93, 021301
\bibitem[Marchesini et al.(2002)]{mar02}
Marchesini, D., D'Onghia, E., Chincarini, G., Firmani, C.,
Conconi, P., Molinari, E., Zacchei, A. 2002, \apj, 575, 801
\bibitem[Meneghetti et al.(2001)]{men01}
Meneghetti, M., Yoshida, N., Bartelmann, M., Moscardini, L.,
Springel, V., Tormen, G., \& White, S. D. M. 2001, \mnras, 325, 435
\bibitem[Mo \& Mao(2000)]{mo00}
Mo, H. J., \& Mao, S. 2000, \mnras, 318, 163
\bibitem[Nipoti et al.(2004)]{nip04}
Nipoti, C., Treu, T., Ciotti, L., Stiavelli, M. 2004, \mnras, 355, 1119
\bibitem[Navarro, Frenk \& White(1996)]{nav96}
Navarro, J. F., Frenk, C. S., \& White, S. D. M. 1996, \apj, 462, 563
\bibitem[Navarro, Frenk \& White(1997)]{nav97}
Navarro, J. F., Frenk, C. S., \& White, S. D. M. 1997, \apj, 490, 493 
\bibitem[Ostriker(2000)]{ost00}
Ostriker, J. P. 2002, \prl, 84, 5258
\bibitem[Qin \& Wu(2001)]{qin01}
Qin, B., \& Wu, X. P. 2001, \prl, 87, 061301
\bibitem[Quinlan(1996)]{qui96}
Quinlan, G. D. 1996, NewA, 1, 255
\bibitem[Romanishin, Strom \& Strom(1982)]{rom82}
Romanishin, W., Strom, S. E., \& Strom, K. M. 1982, \apj, 258, 77
\bibitem[S\'anchez-Salcedo(2003)]{sal03}
S\'anchez-Salcedo, F. J. 2003, \apj, 591, L107
\bibitem[S\'anchez-Salcedo \& Reyes-Ruiz(2004)]{san04}
S\'anchez-Salcedo, F. J., \& Reyes-Ruiz, M. 2004, \apj, 607, 247 
\bibitem[Sancisi(2003)]{san03}
Sancisi, R. 2003 in Dark Matter in Galaxies, IAU Symp.~220,
eds.~S. D. Ryder, D. J. Pisano, M. A. Walker, \& K. C. Freeman, 
San Francisco: Astronomical Society of the Pacific, 233
\bibitem[Simon et al.(2004)]{sim04}
Simon, J. D., Bolatto, A. D., Leroy, A., Blitz, L., \& Gates, E. L.,
\apj, 621, 757
\bibitem[Spergel \& Steinhardt(2000)]{spe00}
Spergel, D. N., \& Steinhardt, P. J. 2000, \prl, 84, 3760
\bibitem[Wandelt et al.(2001)]{wan01}
Wandelt, B. D., Dav\'e, R., Farrar, G. R., McGuirre, P. C., Spergel,
D. N., \& Steinhardt, P. J. 2001, in Sources and Detection of 
Dark Matter and Dark Energy in the Universe, ed.~D.~B.~Cline, (Springer, 
Berlin), 263
\bibitem[Yoshida et al.(2000)]{yos00}
Yoshida, N., Springel, V., White, S. D. M., \& Tormen, G. 2000,
\apj, 544, L87

\end{thebibliography}
\end{document}